\documentclass[prb, preprint, onecolumn, aps, longbibliography]{revtex4-2}
\usepackage{amsmath,amssymb,graphicx}
\usepackage{ulem}
\usepackage{subfigure}
\usepackage{color}
\usepackage{times}
\usepackage{float}
\usepackage{amsfonts}
\usepackage{bm}
\usepackage{mathtools}
\usepackage{hyperref}
\usepackage[utf8]{inputenc}

\DeclareMathOperator{\sgn}{sgn}
\newcommand{\f}{\varphi}
\def\e{\varepsilon}
\def\f{\varphi}
\def\phi{\varphi}
\renewcommand{\Re}{{\rm Re}}

\renewcommand{\d}{{\rm d}}

\newcommand{\w}{\omega}

\newcommand{\ii}{{\rm i}}

\renewcommand{\Re}{{\rm Re}}

\begin{document}

%\title{Nonreciprocal Scattering from Gyrotropic Plasmonic Cylinders and Its Application to Diffractive Metasurfaces}
%\title{Nonreciprocal Photonic Response \\from Gyrotropic Rods and Metasurfaces}
\title{Maximum non reciprocity in metasurfaces of gyrotropic rods}

\author{Ioannis Katsantonis}
\email{katsantonis@iesl.forth.gr}
\affiliation{Institute of Electronic Structure and Laser, Foundation of Research and Technology Hellas, GR-71110, Heraklion, Greece}

\author{Constantinos Valagiannopoulos}
\email{valagiannopoulos@ece.ntua.gr}
\affiliation{School of Electrical and Computer Engineering, National Technical University of Athens, GR-15780, Athens, Greece}

\author{Anna C. Tasolamprou}
\email{atasolam@phys.uoa.gr}
\affiliation{Department of Physics, National and Kapodistrian University of Athens, GR-15772, Athens, Greece}

\begin{abstract}
Efficiently breaking time-reversal symmetry at the subwavelength scale remains a cornerstone challenge for advanced electromagnetic wave manipulation. This work presents a rigorous analytical framework, based on cylindrical wave expansion, to investigate and optimize the nonreciprocal scattering of transverse electric waves by magnetically biased plasmonic rods. An intuitive metric is introduced to quantify the breaking of time-reversal symmetry via the asymmetric lifting of degeneracy between azimuthal modes of opposite angular momentum, hosted by the gyrotropic particles. Leveraging this metric, a comprehensive mapping of the multiparametric space of operational frequency, cyclotron frequency, and cylinder optical size isolates regimes of maximum nonreciprocity. A detailed multipolar decomposition reveals that this extreme behavior stems from the phase-matched asymmetric excitation and interference of localized electric and magnetic dipole modes. Moving from individual, isolated meta-atoms to collective photonic systems, the optimized cylinders are arranged into a periodic grating. Under oblique incidence, the combination of geometric asymmetry and magnetic mode splitting, forces the metasurface to transmit light in a totally different way when excited by opposite sides. The reported findings and design principles offer a versatile blueprint for the development of dynamically tunable flat-optics isolators, directional transceivers, and advanced wavefront routers.

\end{abstract}

\maketitle

\section{Introduction}
\label{sec:SectionI}

In the modern engineering of photonic synthetic matter, a key goal is to control how electromagnetic waves propagate along specific directions. Most passive structures made from conventional media follow the Lorentz reciprocity principle \cite{Asadchy2020}, which imposes natural symmetries on how waves are transmitted, reflected, or scattered. Breaking these interchangeability rules unlocks several advanced functionalities. These include optical isolation \cite{Davoyan:13,jalas}, one-way wave propagation \cite{MannSounasAlu,PhysRevApplied.16.044011}, asymmetric diffraction \cite{SheikhAnsariIyerGholipour+2023+2639+2667,dnp4-z1xk}, and nonreciprocal waveform steering \cite{Tan_sc_rep,Jiaruo}. Ultimately, these capabilities are crucial for next-generation communication, sensing, and information processing \cite{Ataloglou,Monticone_Alu}. Among the various approaches to achieving nonreciprocal behavior, perhaps the most conceptually simple is to magnetically bias an ordinary plasmonic substance; this enriches the host with gyrotropic properties that break time-reversal symmetry, leading to directionally dependent wave responses \cite{Joan_S,Ju:17,Giessen_NatCom}. Consequently, recent research has increasingly focused on leveraging magneto-optical effects in photonic crystals and metallic nanostructures to build compact nonreciprocal devices operating from the microwave to the optical bands \cite{Fang2011,Prudencio2016,Mirko,8979321}.

At the subwavelength scale, the electromagnetic response of individual particles under magnetic bias differs from reciprocal resonators. Early foundations outlined magneto-optical scattering in multi-layer magnetic and dielectric films \cite{Smith1965}, enabling the use of gyrotropic and anisotropic media for mode conversion in optical waveguides \cite{Wang1972}. For three-dimensional geometries, expanding Mie theory to a gyrotropic sphere provided exact analytical solutions for scattering and absorption \cite{Ford1978}. This framework was broadened through potential formulations of rotationally symmetric anisotropic spheres \cite{Qiu2007} and T-matrix characterizations \cite{Li2012} detailing the hybrid effects of concurrent electric and magnetic gyrotropy \cite{Zouros2021}. Beyond optics, gyrotropic structures induce spin orientation and spin currents in quantum wells via spin-orbit coupling \cite{Tarasenko2006} and drive a gyrotropic magnetic effect tied to the Fermi surface \cite{Zhong2016}. Furthermore, noncentrosymmetric metals exhibit a distinct ac gyrotropic Hall effect governed by Berry curvature \cite{Konig2019}. In device applications, multiple-scattering models for uniaxial gyrotropic scatterers facilitate magneto-optical circulators operating under uniform external fields \cite{Smigaj2010}. Ultimately, Zeeman gyrotropic scatterers yield extreme anomalies, including resonance splitting, pattern rotation, and non-radiating embedded eigenstates that bypass reciprocal limitations \cite{Valagiannopoulos2018}.

To bridge the gap between individual subwavelength scatterers and macroscopic wave engineering, recent paradigms have turned to collective meta-atom configurations arranged as gyrotropic metasurfaces and metamaterials. Early milestones demonstrated that strong polarization rotatory power surpassing natural quartz becomes feasible via the electromagnetic coupling of twisted bilayered planar metal patterns \cite{Rogacheva2006} while one-dimensional gyrotropic gratings were shown to magnify diffracted magneto-optical rotations several times \cite{Lu2008}. To bypass bulky magnetic fields, modern architectures achieve truely nonreciprocal gyrotropy without permanent magnets by using electrostatic voltage biases on traveling wave ring resonators \cite{Sounas2013} or embedding active unidirectional electronic circuits directly into the bulk media \cite{Wang2012}. Concurrently, researchers have broken symmetries by pairing resonant antennas on gyromagnetic substrates to trigger sharp Fano interferences \cite{Mousavi2014}. Beyond pure photonics, this polarization rotating wave loss has been leveraged to design bioinspired high entropy alloy metamaterials capable of broad multi wave absorption across microwave, ultrasonic, and water wave regimes \cite{Huang2020}. Moreover, the synergy between magneto-optical effects and the spatial rotation of resonant cylinders has given rise to subwavelength nonreciprocal metasurfaces that unlock unidirectional wavefront manipulation and nonreciprocal spin selection \cite{Pan2024}.

In this work, we investigate the nonreciprocal electromagnetic response of infinite plasmonic cylinders magnetically biased along their axis. To describe the scattering of transverse electric waves by individual gyrotropic rods, we develop a rigorous analytical formulation based on cylindrical wave expansion. By introducing an intuitive metric that quantifies the asymmetry between azimuthal scattering modes of bipolar order, we pinpoint the exact operating regimes that maximize this nonreciprocal behavior. We visually demonstrate these features through the rotation of radiation patterns, which dynamically flip upon reversing the magnetic bias sign. To unveil the underlying physics, we utilize a multipolar decomposition of the scattering response. Leveraging these optimized gyrotropic cylinders, we arrange them into an infinite periodic array to construct a metasurface, analyzing its diffraction properties to reveal direction-dependent transmission and reflection under opposite oblique incidences. Notably, peak nonreciprocity emerges at specific combinations of incident angles and grating periods that coincide with the onset of new diffraction orders. Finally, we spectrally investigate exemplary cases hosting anomalous transmission and reflection, validating the overall nonreciprocal response through full-wave numerical simulations. These findings hold significant promise for advancing next-generation nonreciprocal optical devices, including compact isolators, directional routers, and magnetically tunable metasurfaces for advanced beam-steering applications.

\begin{figure}[h!]
\centering
%\subfigure[]{\includegraphics[height=5.9cm]{Fig1a}
%   \label{fig:Fig1a}}
%\subfigure[]{\includegraphics[height=5.9cm]{Fig1b}
%   \label{fig:Fig1b}}
\includegraphics[width=8.5cm]{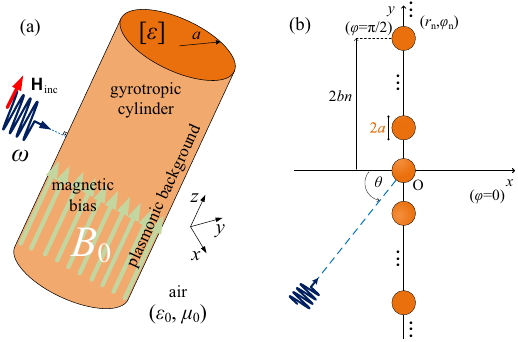}
\caption{(a) The building block of the considered setup. An infinitely long plasmonic cylinder biased by an axial magnetic field to inherit it with gyrotropic properties. TE$_z$ fields are assumed, able to experience the developed anisotropy. (b) Infinite cylinders positioned along vertical $y$ axis to form a metasurface that gets obliquely excited by a plane TE$_z$ wave whose direction of propagation forms an angle $\theta$ with the horizontal $x$ axis.}
\label{fig:Fig1}
\end{figure}

\section{Individual Gyrotropic Rod}
\label{sec:Metaatom}

\subsection{Definitions and Approximations}
\label{sec:ConfDef}
The building unit of our configuration is depicted in Fig.~\ref{fig:Fig1}(a) where the  Cartesian coordinate system $(x,y,z)$ and the equivalent cylindrical $(r,\f,z)$, are also defined. It consists of an infinite plasmonic rod of relative permittivity $\e_p$ and radius $a$ gets biased by an axial magnetic DC field of magnitude $B_0$. When an axial static magnetic field is applied to a plasmonic (free-electron) medium, the electron gas is described by the magnetized Drude plasma model. The motion of the conduction electrons is affected by the Lorentz force, and the electron dynamics are no longer isotropic in the plane perpendicular to the bias. As a result, the permittivity becomes anisotropic and non-reciprocal, and it is described by a gyrotropic permittivity tensor instead of a scalar dielectric constant. The gyrotropic material in the Cartesian coordinate system, it is characterized by the following relative permittivity tensor: 
\begin{eqnarray}
\overline{\overline{\boldsymbol{\e}}}=\left[\begin{array}{ccc}
           \e_t    & -\ii \e_c & 0 \\ 
          \ii \e_c &  \e_t    & 0 \\
					   0     &   0      & \e_p 
		 \end{array}\right].
\label{PermittivityMatrix}
\end{eqnarray}
The permittivity $\e_t$ is the transverse one in the 
$x–y$ plane, $\e_c$ is the gyrotropic (magneto-optical) coefficient, responsible for nonreciprocity,
$\e_p$ is the axial permittivity along the magnetic bias direction.

As far as the dispersion of the plasmonic material is concerned, we assume a simple Drude model:
\begin{eqnarray}
\e_p(\w)=1-(\w_p/\w)^2,
\label{PlasmonicPermittivity}
\end{eqnarray}
while the respective quantities for the magnetized Drude plasma  appeared in \eqref{PermittivityMatrix} read \cite{BittencourtBook}:
\begin{eqnarray}
\e_t(\w)=1-\frac{\omega_p^2}{\omega^2-\omega_c^2}~~~,~~~
\e_c(\w)=\frac{\omega_p^2\omega_c}{\omega\left(\omega^2-\omega_c^2\right)},
\label{GyrotropicPermittivities}
\end{eqnarray}
where $\omega_p$ is the plasma frequency and $\omega_c$ the cyclotron frequency proportional to the magnetic bias $B_0$. We observe that all the permittivities $\{\e_p,\e_t,\e_c\}$ of \eqref{PlasmonicPermittivity},\eqref{GyrotropicPermittivities} can be controlled by two variables: the normalized operational frequency $\omega/\omega_p$ and the cyclotron frequency $\omega_c/\omega_p$ divided by the plasma frequency $\omega_p$ of that material. Harmonic time-dependence of the form $\exp(+\ii\w t)$ is suppressed throughout.

In the whole analysis, TE$_z$ waves are  considered, namely, magnetic fields dependent on $(x,y)$ or $(r,\f)$ and parallel to $z$ axis, so that signal activates the anisotropy of the rod \cite{Valagiannopoulos2007}. The wave equation in the gyrotropic medium is rigorously solved by using the Cartesian to cylindrical coordinate transformation for the tensor \eqref{PermittivityMatrix} with eigenfuntions $\{J_u(k_0 r \sqrt{\e_g})e^{\ii u \f}\}$ for $u\in\mathbb{Z}$.  $J_u$ is  the Bessel function of order $u$, $k_0=2\pi/\lambda=\w \sqrt{\e_0\mu_0}$ is the wavenumber into free space while the effective permittivity $\e_g$ reads: 
\begin{eqnarray}
\e_g=\frac{\e_t^2-\e_c^2}{\e_t}.
\label{EffectivePermittivity}
\end{eqnarray}
Once the cylinder is smaller than the operational wavelength $\lambda$, one can perform the dipolar approximation for the magnetic field $H_g(r,\f)$ into its volume by keeping only the orders $u=0$ and $u=\pm 1$ from the infinite sum of eigensolutions of the partial wave formalism. Indeed, regardless of the signal outside the rod, if $a<\lambda$, it can be written as:
\begin{eqnarray}
H_g(r,\f)\cong C_0 J_0(k_g r)~~~~~~~ \nonumber\\
+\left[C_{1c}\cos\f+C_{1s}\sin\f \right]J_1(k_g r).
\label{PlasmonicMagneticField}
\end{eqnarray} 
The symbol $J_u$ is used for the Bessel function of order $u\in\mathbb{Z}$ while $k_g=k_0\sqrt{\e_g}$.

\subsection{Maximally Nonreciprocal Cylinders}
\label{sec:MaxNonRecRod}
It is well-known \cite{SounasPaper} that the magnetic bias $B_0$ renders the rods nonreciprocal, which is a highly sought-after feature of materials allowing for control over the electromagnetic waves, in a way that goes far beyond ordinary filtering \cite{OurSRPaper}. Therefore, it is meaningful to examine the conditions under which the meta-atoms behave in a maximally nonreciprocal way \cite{MyJ57}. Let us consider a free-standing particle like that depicted in Fig.~\ref{fig:Fig1}(a), under a plane wave TE$_z$ illumination similar to that of Fig.~\ref{fig:Fig1}(b), namely, $H_{\rm inc}=e^{-\ii k_0(x \cos\theta+y \sin\theta)}$. The scattered ($z$-polarized) magnetic field reads: $H_{\rm scat}=\sum_{u=-\infty}^{+\infty}S'_u H_u(k_0 r) e^{\ii u (\f-\theta)}$, where the coefficients $S'_u$ for $u\in\mathbb{Z}$ are obtained analytically. As previously stated, an optically thin cylinder ($k_0a\ll 1$) is well-described by keeping up to the dipolar partial waves terms ($u=0,\pm 1$), in the aforementioned sum. In such a scenario, how different is the $u=+1$ term ($S'_1$) from the opposite of the $u=-1$ term ($S'_{-1}$) is proportional to how nonreciprocal is the regarded particle; indeed, it vanishes by setting the cyclotron frequency equal to zero ($\w_c=0$). For this reason, let us define as a good metric of nonreciprocity, the quantity:
\begin{eqnarray}
\Delta S'=\left|S'_1+S'_{-1}\right|.
\label{NonRecIndex}
\end{eqnarray}
With use of approximations of Bessel and Hankel functions with small arguments ($a\ll \lambda$) \cite{AbramowitzStegun}, it is found that \eqref{NonRecIndex} blows up for:
\begin{eqnarray}
(\w/\w_p)^2+\frac{1}{4 (\w/\w_p)^2}-1=(\w_c/\w_p)^2,
\label{NonRecCondition}
\end{eqnarray}
which also defines our operational regimes.   
 
\begin{figure}[ht!]
\centering
\includegraphics[width=8.5cm]{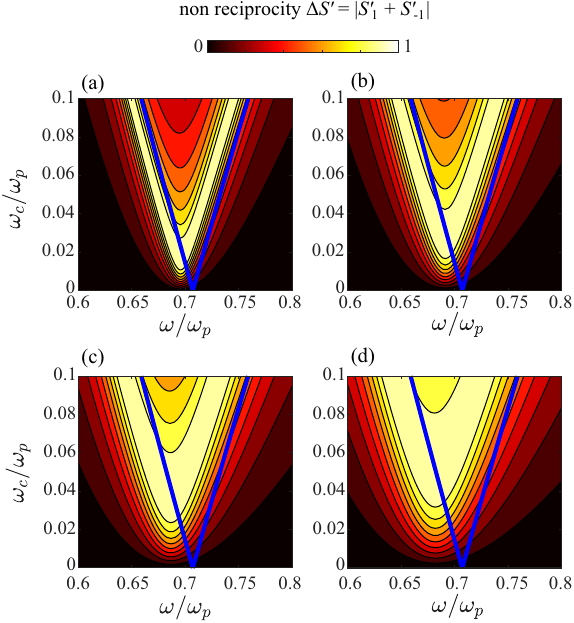}
%\subfigure[]{\includegraphics[width=4.2cm]{Fig2b}
%   \label{fig:Fig2b}}\\
%\subfigure[]{\includegraphics[width=4.2cm]{Fig2c}
%   \label{fig:Fig2c}}
%\subfigure[]{\includegraphics[width=4.2cm]{Fig2d}
%   \label{fig:Fig2d}}
\caption{The nonreciprocity metric $\Delta S'$ defined in \eqref{NonRecIndex}, as function of normalized operational frequency $\w/\w_p$ and normalized cyclotron frequency $\w_c/\w_p$ for cylinders with various optical radii. (a) $a/\lambda_p=0.04$, (b) $a/\lambda_p=0.05$, (c) $a/\lambda_p=0.06$, (d) $a/\lambda_p=0.07$. The blue lines correspond to the approximate conditions \eqref{NonRecCondition}.}
\label{fig:Fig2}
\end{figure}

In Fig.~\ref{fig:Fig2}, we show that nonreciprocity index from \eqref{NonRecIndex} in contour plots with horizontal axis indicating the normalized oscillating frequency $\w/\w_p$ and vertical axis measuring the normalized cyclotron frequency $\w_c/\w_p$, for several optical sizes of the cylinder $a/\lambda_p$. The symbol $\lambda_p$ is used for the free-space wavelength corresponding to plasma frequency $\w_p$. In addition, blue lines represent the parametric loci dictated by the approximate resonant conditions \eqref{NonRecCondition}. In Fig.~\ref{fig:Fig2}(a), we regard the smallest meta-atom ($a=0.04\lambda_p$) and observe that for low $\w_c$ a single maximum emerges around $\w\cong 0.7\w_p$ while, for increasing $\w_c$, a Zeeman effect occurs happening when an external magnetic field results in the splitting of spectral lines \cite{ZeemanEffect}. That magnetic bias gives “rotational preference” to the system and makes the moving electrons behave differently depending on their sense of direction. In this way, two loci of high $\Delta S'$ are formulated that are well-predicted by the approximate condition \eqref{NonRecCondition}. 

%\begin{figure}[ht!]
%\centering
%\subfigure[]{\includegraphics[width=4.2cm]{Fig2a}
 %  \label{fig:Fig2a}}
%\subfigure[]{\includegraphics[width=4.2cm]{Fig2b}
%   \label{fig:Fig2b}}\\
%\subfigure[]{\includegraphics[width=4.2cm]{Fig2c}
%   \label{fig:Fig2c}}
%\subfigure[]{\includegraphics[width=4.2cm]{Fig2d}
%   \label{fig:Fig2d}}
%\caption{The nonreciprocity metric $\Delta S'$ defined in \eqref{NonRecIndex}, as function of normalized operational frequency $\w/\w_p$ and normalized cyclotron frequency $\w_c/\w_p$ for cylinders with various optical radii. (a) $a/\lambda_p=0.04$, (b) $a/\lambda_p=0.05$, (c) $a/\lambda_p=0.06$, (d) $a/\lambda_p=0.07$. The blue lines correspond to the approximate conditions \eqref{NonRecCondition}.}
%\label{fig:Figs2}
%\end{figure}

In Fig.~\ref{fig:Fig2}(b), we regard a larger rod ($a=0.05\lambda_p$) and the two parametric strips of substantial $\Delta S'$ become wider while appear at higher cyclotron frequencies. As far as the estimation \eqref{NonRecCondition} is concerned, it can also give a good initial guess for the optimal operational regimes of maximum $\Delta S'$. The same trend continues in Fig. \ref{fig:Fig2}(c) which is an indication that more sizable cylinders require higher magnetic biases $B_0$ to acquire the same degree of nonreciprocity. Naturally, in Fig. \ref{fig:Fig2}(d), where the thickest design ($a=0.07\lambda_p$) is considered, the split of the spectral lines do not emerge within the parametric window $(\w/\w_p,\w_c/\w_p)$ used for all the selected rods.  

It should be stressed that we do not investigate thinner cylinders ($a<0.04\lambda_p$) because their response will be mainly omni-directional and, thus, nonreciprocity is suppressed. Similarly, we avoid examining bigger designs ($a>0.07\lambda_p$) since higher-order azimuthal harmonics will emerge and the nonreciprocity will become less pronounced. When it comes to cyclotron frequency, we keep it fixed at $\w_c=0.05 \w_p$ in our examples; such a choice does not seem to be restrictive in terms of the obtained conclusions. In fact, the maximal value of $\Delta S'$ from \eqref{NonRecIndex} is equal to one and is achieved when one of the two dipolar coefficients $S'_{\pm 1}$ gets almost equal to unity and the other vanishes. This can happen either for $|S'_{-1}|^2\cong 1$ while $|S'_{+1}|^2\ll 1$, meaning that the $\exp(-\ii \f)$ is the dominant azimuthal terms, or for $|S'_{+1}|^2\cong 1$ while $|S'_{-1}|^2\ll 1$, leading to waves with $\exp(+\ii \f)$ angular variation. The Zeeman split of Fig. \ref{fig:Fig2} hosts both solutions, one at each branch. 

In every single map of Fig.~\ref{fig:Fig2} we optimize the metric $\Delta S'$ to ensure that it takes unitary value; in this sense, we obtain maximally nonreciprocal particles without sweeping the parameter $\w_c/\w_p$. If two solutions are available (like in Figs \ref{fig:Fig2}(a) and \ref{fig:Fig2}(b)), we select the one emerging at higher oscillating frequency $\w/\w_p$ to work with the optically bigger rod; these designs support waves with $\exp(-\ii \f)$ variation. Once a single maximum appears (like in Figs \ref{fig:Fig2}(c) and \ref{fig:Fig2}(b)) we, apparently, pick it. In this way, we conclude to four cylinders of various sizes, operated at different frequencies that all support maximally nonreciprocal scattering ($\Delta S' \rightarrow 1$).  

\begin{figure}[ht!]
\centering
\includegraphics[width=6.0cm]{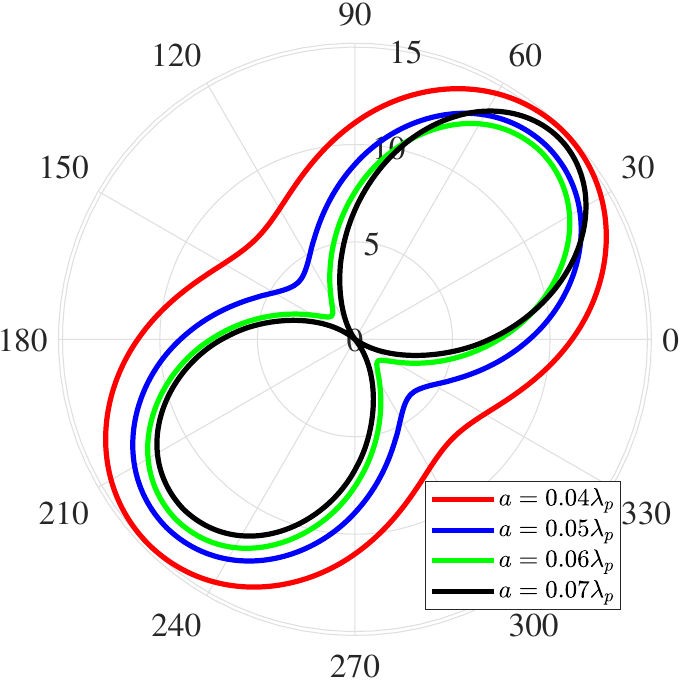}
\caption{The scattering response $s(\f,0)=s(\f,\pi)$ of selected nonreciprocal designs taken from Fig.~\ref{fig:Fig2} when they are excited along the horizontal axis ($\theta=0$ or $\theta=\pi$). Cyclotron frequency: $\w_c/\w_p=0.05$. Operational frequency $\w/\w_p$ is picked, different for each curve, from the maps of Fig.~\ref{fig:Fig2} to lead to maximum $\Delta S'$ from \eqref{NonRecIndex}.}
\label{fig:Fig3}
\end{figure}

In Fig.~\ref{fig:Fig3}, we represent the responses of these four optimized individual rods in the far region. Note that the scattered power in the far field is computed with help from the rigorous formula: $s(\f,\theta)=\frac{2}{k_0 a}\left|\sum_{u=-\infty}^{+\infty}S'_u \ii^u e^{\ii u (\f-\theta)}\right|^2$, incorporating all possible harmonics \cite{Valagiannopoulos2011}. Note that if one takes the average value of $s(\f,\theta)$ around the circle, one obtains the total scattered power $P_{\rm scat}$ over the incident power $P_{\rm inc}$ passing through the $2a$ aperture of the cylinder if the latter is conditionally removed, namely: $P_{\rm scat}/P_{\rm inc}=\frac{1}{2\pi}\int_0^{2\pi}s(\f,\theta)\d\f$. Importantly, for all the four optimal cylinders the scattering power density is the same at the arbitrary direction $\f$ either we illuminate the structure by positively or negatively propagating waves. Therefore, we obtain $s(\f,0)=s(\f,\pi)$ because $\Delta S' \cong 1$ and the asymmetry with respect to horizontal ($x$) axis is an indication of how nonreciprocally our rods behave. By inspection of Fig.~\ref{fig:Fig3}, we notice that all the polar curves look symmetric with respect to rotated axes of specific angle $\f\ne 0$, namely are hugely asymmetric with respect to $x$ direction, which demonstrates the detected nonreciprocity under excitation along the horizontal axis ($\f=0$). As mentioned above, maximum nonreciprocity is also indicated by the fact that the same (asymmetric) curves are obtained if we assume a plane wave traveling along the negative $x$ semi-axis ($\theta=\pi$). 

\begin{figure}[ht!]
\centering
%\subfigure[]{\includegraphics[width=4.2cm]{Fig4a}
  % \label{fig:Fig4a}}
%\subfigure[]{\includegraphics[width=4.2cm]{Fig4b}
 %  \label{fig:Fig4b}}\\
%\subfigure[]{\includegraphics[width=4.2cm]{Fig4c}
 %  \label{fig:Fig4c}}
%\subfigure[]{\includegraphics[width=4.2cm]{Fig4d}
%   \label{fig:Fig4d}}
\includegraphics[width=8.3cm]{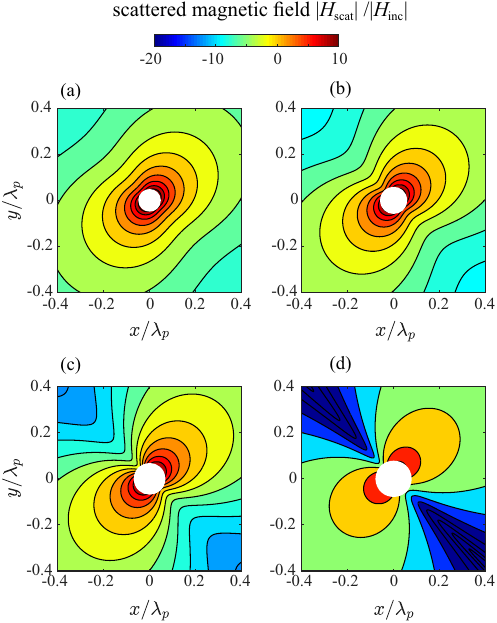}
\caption{The magnitude of the scattering magnetic field $|H_{\rm scat}|$ normalized by $|H_{\rm inc}|$ (in dB) externally to each of the optimal designs of Fig.~\ref{fig:Fig3}, when excited by a horizontally propagating plane wave ($\theta=0,\pi$) of unitary magnetic field. (a) $a/\lambda_p=0.04$ ($\w\cong 0.719\w_p$), (b) $a/\lambda_p=0.05$ ($\w\cong 0.712\w_p$), (c) $a/\lambda_p=0.06$ ($\w\cong 0.702\w_p$), (d) $a/\lambda_p=0.07$ ($\w\cong 0.682\w_p$). Cyclotron frequency: $\w_c/\w_p=0.05$. The cross sections of the cylinders appear in white color since we are interested for points with $r>a$.}
\label{fig:Fig4}
\end{figure}

It is also noteworthy that the scattering of all the particles is strong since the value of $s(\f,\theta)$ says how many times larger is the power density of the scattered wave compared to the incident one; in all the considered examples, we record peaks close to $s\cong 15$. In addition, the larger the cylinder is, the more suppressed the omnidirectional component gets; indeed, for the most sizable sample ($a=0.07\lambda_p$) only dipolar terms (in fact, the $u=-1$ one) are practically present. It should be also noticed that no other from the harmonics of higher order $|u|\ge 1$ seems having significant contribution.

The distribution of the scattered magnetic field $|H_{\rm scat}|$ in the near region for every single of the four optimized rods, is depicted in Fig.~\ref{fig:Fig4}. It is clear that the same rotations as those observed in the far field (Fig.~\ref{fig:Fig3}) are recorded in the spatial vicinity of each scatterer. Note also that, even in the near field, the response of our four optimal design are identical either we excite the cylinder from $\theta=0$ or from $\theta=\pi$.

\begin{figure}[ht!]
\centering
%\subfigure[]{\includegraphics[width=4.2cm]{Fig45a}
%   \label{fig:Fig45a}}
%\subfigure[]{\includegraphics[width=4.2cm]{Fig45b}
%   \label{fig:Fig45b}}
\includegraphics[width=8.3cm]{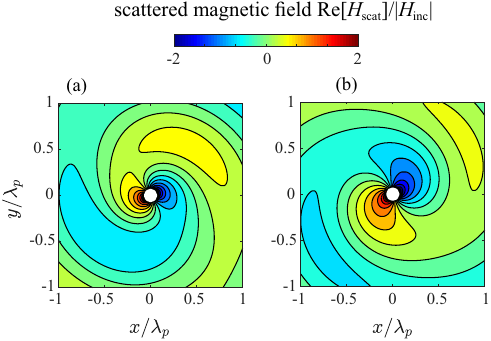}
\caption{The real part of the scattering magnetic field $\Re[H_{\rm scat}]$ normalized by $|H_{\rm inc}|$ (in dB) externally to the two optimal designs of Fig.~\ref{fig:Fig2}(b) ($a=0.05\lambda_p$), when excited by a horizontally propagating plane wave ($\theta=0,\pi$) of unitary magnetic field. (a) $\w \cong 0.671 \w_p$, (b) $\w \cong 0.712 \w_p$. Cyclotron frequency: $\w_c/\w_p=0.05$.}
\label{fig:Fig45}
\end{figure}

One may wonder what is the difference between the selected optimal designs from Fig.~\ref{fig:Fig2} and the ones with the same nonreciprocity maximum score \eqref{NonRecIndex} $\Delta S'=1$ but appearing at lower frequencies $\w/\w_p$ (with only $u=+1$ harmonic present). Indeed, in Figs \ref{fig:Fig2}(a) and \ref{fig:Fig2}(b), there are frequencies at which $\Delta S'=1$ lying into the left branch. In Fig.~\ref{fig:Fig45}, we represent the near-field scattering response $\Re[H_{\rm scat}(x,y)]$ for the two optimal designs of Fig.~\ref{fig:Fig2}(b) ($a=0.05\lambda_p$) yielding a maximal nonreciprocity ($\Delta S'=1$). In Fig.~\ref{fig:Fig45}(a), we consider the cylinder operated at $\w\cong 0.671 \w_p$ and realize that the scattering pattern is rotated while the resonant response  corresponds to a circularly polarized induced dipole with counter-clockwise sense of rotation ($\exp(+\ii\f)$). In Fig.~\ref{fig:Fig45}(b), we shift the oscillation frequency to $\w\cong 0.712 \w_p$, namely we obtain the case of Fig.~\ref{fig:Fig4}(b). One directly observes a similar helical distribution of $\Re[H_{\rm scat}(x,y)]$ but possessing the opposite sense of rotation from that of Fig.~\ref{fig:Fig45}(a) ($\exp(-\ii\f)$). Once again, such a difference in the angular momentum \cite{EskinPaper} is attributed to the fact that one of the two coefficients $|S'_{\pm 1}|$ becomes much larger than the other ($|S'_{-1}|\gg |S'_{+1}|$ in Fig.~\ref{fig:Fig45}(b)). It is, finally, remarkable that the magnitude of the magnetic scattered field both in the near (Fig. \ref{fig:Fig4}(b) and in the far (Fig. \ref{fig:Fig3}) region for the two oscillation frequencies leading to maximal nonreciprocity \eqref{NonRecIndex} differ only slightly.

\subsection{Multipolar Expansion}
\label{sec:MultipolAna}
To further unveil the physical mechanisms responsible for the nonreciprocal scattering by the gyrotropic rods, we analyze their electromagnetic response by using a multipolar decomposition. This approach allows us to identify the relative contributions of the fundamental scattering channels and to clarify the role of electric and magnetic dipolar modes in the observed asymmetric radiation patterns. Subsequently, from the magnetic field vector, $\textbf{H}=\hat{\textbf{z}}H(x,y)=\hat{\textbf{z}}H(r,\f)$, we calculate the polarization current as: $\textbf{J}=\left(\overline{\overline{\boldsymbol{\e}}}\right)^{-1}\cdot\left(\textbf{I}-\overline{\overline{\boldsymbol{\e}}}\right)\cdot\left(\nabla\times\textbf{H}\right)$. The symbol $\textbf{I}$ is used for the $3\times 3$ identity matrix. From the Cartesian components of the current density $\textbf{J}=J_x\hat{\textbf{x}}+J_y\hat{\textbf{y}}+J_z\hat{\textbf{z}}$, we quantify the role of the electric dipole moment vector (ED) $\textbf{p}=p_x\hat{\textbf{x}}+p_y\hat{\textbf{y}}+p_z\hat{\textbf{z}}$, the magnetic dipole moment vector (MD) $\textbf{m}=m_x\hat{\textbf{x}}+m_y\hat{\textbf{y}}+m_z\hat{\textbf{z}}$ as follows \cite{PhysRevB.89.205112, IVAN,PhysRevApplied.15.014043}, \cite{MinimalScattering}.
\begin{subequations}
\label{EMMoments}
\begin{eqnarray}
\textbf{p} & = & \frac{1}{\ii \w}\int_{\textrm{(A)}}\textbf{J}~\d A, \\
\textbf{m} & = & \frac{\sqrt{\e_0\mu_0}}{2}\int_{\textrm{(A)}}\left(\textbf{r}\times\textbf{J}\right)~\d A,
\end{eqnarray}
\end{subequations}
where $\d A=\d x\d y$ is the differential area over the cross section of the cylinder $(r<a)$ denoted as $\textrm{(A)}$. The position vector reads $\textbf{r}=x\hat{\textbf{x}}+y\hat{\textbf{y}}+z\hat{\textbf{z}}$. 

As far as the electric quadrapole moment tensor (EQ) $\overline{\overline{\textbf{Q}}}_{\textrm{e}}=[Q_{jl}^{\textrm{e}}]$ and magnetic quadrapole moment tensor (MQ) $\overline{\overline{\textbf{Q}}}_{\textrm{m}}=[Q_{jl}^{\textrm{m}}]$ are concerned, they are expressed in Cartesian coordinates with $j=x,y,z$ and $l=x,y,z$ in the following way:
\begin{subequations}
\label{EMMoments2}
\begin{eqnarray}
Q^{\textrm{e}}_{jl} & = & \frac{\ii}{2\w}\int_{\textrm{(A)}}\left[ 3 (l J_j + j J_l ) -\frac{2}{3}(\textbf{r} \cdot \textbf{J})\delta_{jl} \right]\d A, \\
Q^{\textrm{m}}_{jl} & = & \frac{\sqrt{\e_0\mu_0}}{3}\int_{\textrm{(A)}}\left[l {\left(\textbf{r}\times\textbf{J}\right)}\cdot \hat{\textbf{j}} +j {\left(\textbf{r}\times\textbf{J}\right)}\cdot \hat{\textbf{l}}\right]\d A.
\end{eqnarray}
\end{subequations}
As implied above, the indexes $j$ and $l$ can refer independently to each of the three Cartesian coordinates $\{x,y,z\}$; that rule concerns the directive position vectors $\textbf{j}$ and $\textbf{l}$ as well as the respective unitary ones $\hat{\textbf{j}}$ and $\hat{\textbf{l}}$. To investigate the full multipolar contributions to the response of nonreciprocity of the gyrorod, the scattering efficiency by a single gyrorod is approximated by including the above first four multipole moments; the respective formula is given below \cite{IVAN}:
\begin{eqnarray} 
\label{eq:10}
\frac{P_{\rm scat}}{P_{\rm inc}}
 \cong \frac{\omega^4 \e_0 \mu_0 }{12\pi a} 
 \left[\sum_{j=x,y,z}\left(|p_j|^2+|m_j|^2\right)\right. \nonumber \\+ \left.\frac{\w^2 \e_0\mu_0 }{120} \sum_{j,l=x,y,z} \left(|Q^{\textrm{e}}_{jl}|^2+|Q^{\textrm{m}}_{jl}|^2\right)\right].
\end{eqnarray}
In the considered case, electric dipole (ED) is represented by two complex components: $\{p_x,p_y\}$, magnetic dipole (MD) by only one: $\{m_z\}$, electric quadrupole (EQ) by three complex terms: $\{Q^{\textrm{e}}_{xx},Q^{\textrm{e}}_{yy},Q^{\textrm{e}}_{xy}\}$ and the magnetic quadrupole (MQ) by two: $\{Q^{\textrm{m}}_{yz},Q^{\textrm{m}}_{xz}\}$. It is remarked that all moments \eqref{EMMoments},\eqref{EMMoments2} are defined per unit length of $z$ axis since the system is two-dimensional; the same convention holds for scattered and incident powers as well as the scattering efficiency $P_{\textrm{scat}}/P_{\textrm{inc}}$ in \eqref{eq:10}. 

This decomposition provides a quantitative way to evaluate the relative weight of each multipolar contribution in the scattering process. In the considered subwavelength regime, the response of the gyrotropic rod is expected to be dominated by dipolar terms, while higher-order multipoles are only weakly present. The multipolar analysis is particularly useful for understanding the origin of the nonreciprocal behavior. In particular, the gyrotropic permittivity tensor breaks time-reversal symmetry under magnetic bias and modifies the coupling between the electric and magnetic dipolar modes. As a result, the scattering response becomes asymmetric with respect to the excitation direction, leading to the directional scattering features observed in the following sections. In the next section, the multipolar decomposition is used to analyze the spectral response of the gyrotropic rod and to identify the conditions under which the electric and magnetic dipole contributions overlap, giving rise to enhanced nonreciprocal scattering.

\begin{figure}[ht!]
\centering
%\subfigure[]{\includegraphics[width=4.2cm]{FigXa}
%   \label{fig:FigXa}}
%\subfigure[]{\includegraphics[width=4.2cm]{FigXb}
%   \label{fig:FigXb}}\\
%\subfigure[]{\includegraphics[width=4.2cm]{FigXc}
%   \label{fig:FigXc}}
%\subfigure[]{\includegraphics[width=4.2cm]{FigXd}
%   \label{fig:FigXd}}
   \includegraphics[width=8.4cm]{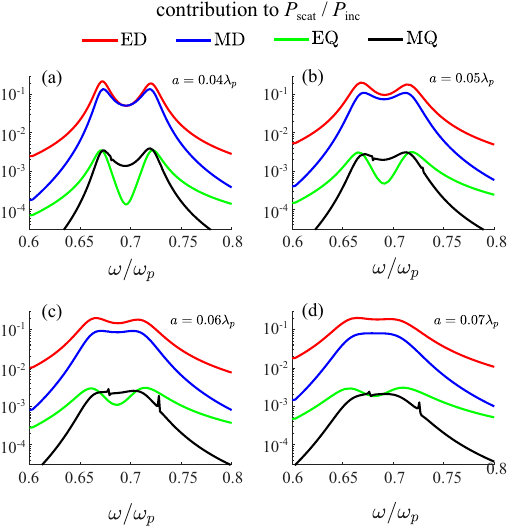}
\caption{The contribution of each of the terms ED (electric dipole), MD (magnetic dipole), electric quadrupole (EQ) and magnetic quadrupole (MQ) according to \eqref{EMMoments},\eqref{EMMoments2}, to the normalized scattered power $P_{\rm scat}/P_{\rm inc}$ in \eqref{eq:10} as function of the normalized operational frequency $\w/\w_p$ for individual particles from Fig. \ref{fig:Fig2} with various optical radii. (a) $a/\lambda_p=0.04$, (b) $a/\lambda_p=0.05$, (c) $a/\lambda_p=0.06$, (d) $a/\lambda_p=0.07$. Cyclotron frequency: $\w_c/\w_p=0.05$.}
\label{fig:FigX}
\end{figure}

To investigate the physical origin of the scattering response, we evaluate the multipolar contributions of the gyrorod using the decomposition described above. In Fig.~\ref{fig:FigX}, we present the spectral distribution of the electric dipole (ED), magnetic dipole (MD), and higher-order multipoles (EQ, DQ) making the total scattering efficiency \eqref{eq:10}. It is observed that the response is predominantly governed by the electric and magnetic dipole modes, while quadrupole contributions remain comparatively weak. This behavior validates the dipolar approximation adopted in \eqref{PlasmonicMagneticField}, which captures the dominant scattering mechanisms of the subwavelength gyrotropic cylinder. Specifically, a strong spectral overlap between the electric and magnetic dipole modes appears near the operational frequency, indicating a balanced magneto-electric interaction regime that plays a crucial role in the asymmetric scattering behavior introduced by the gyrotropic medium.

 Particularly, in Fig. \ref{fig:FigX}(a), we choose the smallest rod ($a=0.04\lambda_p$) and observe that for cyclotron frequency $\omega_c=0.05\omega_p$, a Zeeman splitting effect due to external magnetic field result in two distinct spectral resonances. In Fig. \ref{fig:FigX}(b), we choose a thicker rod ($a=0.05\lambda$) and the two spectral resonances become broader. As the rod radius increases further ($a=0.06\lambda$ and $a=0.07\lambda$), the spectral response becomes progressively flat as shown in Figs  \ref{fig:FigX}(c),\ref{fig:FigX}(d), respectively, indicating a reduction in spectral confinement. Moreover, while for thinner radii the electric and magnetic responses are of comparable strength, for sizable particles the magnetic response becomes significantly weaker compared to the electric one. This also provides further evidence that, as the rod cross section increases, a higher external magnetic field is required in order to achieve the same non-reciprocal effect.

\section{Infinite Gyrotropic Metasurface}
\label{sec:Metasurface}

\subsection{Mathematical Formulation}
\label{sec:MathFormMeta}
The rod analyzed in Section \ref{sec:Metaatom} is the meta-atom in the metasurface depicted in Fig.~\ref{fig:Fig1}(b). Infinitely number of these cylinders being parallel to $z$ axis, are placed along the $y$ axis  with constant distance $2b$ between the centers of two neighboring ones. In Fig.~\ref{fig:Fig1}(b), we define the local coordinate systems $(r_n,\f_n)$ centralized at the axis of each rod, namely, for $n\in\mathbb{Z}$. The metasurface is illuminated by an obliquely incident plane wave of axial magnetic field written as \cite{BalanisBook}: $H_{\rm inc}=e^{-\ii k_0(x \cos\theta+y \sin\theta)}=\sum_{u=-\infty}^{+\infty}\ii^{-u}J_u(k_0r)e^{\ii u(\f-\theta)}$ with unitary amplitude $1~\textrm{A/m}$.

After performing the usual dipolar approximation as in \eqref{PlasmonicMagneticField}, the scattering field by the whole metasurface takes readily the form:
\begin{eqnarray}
H_{\rm scat}(r,\f) \cong \sum_{n=-\infty}^{+\infty}e^{-2\ii k_0bn\sin\theta}\left[S_0 H_0(k_0 r_n)\right.\nonumber\\
+\left.S_{1c}H_1(k_0 r_n)\cos\f_n+S_{1s}H_1(k_0 r_n)\sin\f_n\right],
\label{ScatteringFieldLocalPolar}
\end{eqnarray} 
where $\{S_0,S_{1c},S_{1s}\}$ are the complex coefficients expressing the response of the central meta-atom. Indeed, due the infinite size of the structure and the infinite wavefront of the excitation, the scattering from the arbitrary $n$-th particle will be spatially identical to any other's with a phase difference $e^{-2\ii k_0bn\sin\theta}$ dictated by its position around axis $(x,y)=(0,2bn)$. Hence, the quantities $\{S_0,S_{1c},S_{1s}\}$ are not functions of $n$ and describe the interaction of the central ($n=0$) wire with the impressed field.

By employing the addition theorem for circular cylindrical waves \cite{StrattonBook}, we can find a formula for the scattering field expressed in the global cylindrical coordinate system $(r,\f,z)$, associated with the central particle:
\begin{eqnarray}
H_{\rm scat}(r,\f) \cong (S_0\Sigma_0-S_{1s}\Sigma_1)J_0(k_0r)+S_0 H_0(k_0 r)                            \nonumber\\
        +S_{1c}\left[(\Sigma_0+\Sigma_2)J_1(k_0r)+H_1(k_0 r)\right]\cos\f                      \nonumber\\
	            +\left\{\begin{array}{c}\left[2S_0\Sigma_1+S_{1s}(\Sigma_0-\Sigma_2))\right]J_1(k_0r)\\
				+S_{1s} H_1(k_0 r)\end{array}\right\}\sin\f.
\label{ScatteringFieldGlobalPolar}
\end{eqnarray}
The symbols $\Sigma_0, \Sigma_1, \Sigma_2$ are used for infinite sums involving Hankel functions $H_u$ with $u\in\mathbb{Z}$ defined as:
\begin{subequations}
\label{AuxiliarySums}
\begin{eqnarray}
\Sigma_0 & = & \sum_{0\ne n=-\infty}^{+\infty}e^{-2\ii k_0bn\sin\theta}H_0(2k_0b|n|), \\
\Sigma_1 & = & \sum_{0\ne n=-\infty}^{+\infty}e^{-2\ii k_0bn\sin\theta}H_1(2k_0b|n|)\sgn(n), \\
\Sigma_2 & = & \sum_{0\ne n=-\infty}^{+\infty}e^{-2\ii k_0bn\sin\theta}H_2(2k_0b|n|).
\end{eqnarray}
\end{subequations}
The derivation of \eqref{ScatteringFieldGlobalPolar} and the numerical evaluation of \eqref{AuxiliarySums} have been thoroughly described \cite{TagayPaper}. 

The next step is to impose the necessary boundary conditions just around the $n=0$ cylinder at $r=a$, since the rest of particles interact in a trivially different fashion from the central one (due to the aforementioned phase difference $e^{-2\ii k_0bn\sin\theta}$). Indeed, we have only six complex unknowns $\{C_0,C_{1c},C_{1s},S_0,S_{1c},S_{1s}\}$ and two boundary conditions at $r=a$: (i) Dirichlet: $H_{\rm inc}+H_{\rm scat}=H_g$ and (ii) Neumann: $\frac{\partial H_{\rm inc}}{\partial r}+\frac{\partial H_{\rm scat}}{\partial r}=\frac{1}{\e_g}\frac{\partial H_g}{\partial r}-\ii \frac{\e_c/r}{\e_t^2-\e_c^2}\frac{\partial H_g}{\partial \f}$. The provided constraints are sufficient since each boundary condition will contribute with three scalar equations (one independent from $\f$, another proportional to $\cos\f$ and the last proportional to $\sin\f$). Note that dipole approximation is also performed in the incident wave since a small particle cannot experience significant azimuthal variation of the background field. 

After determining the unknowns $\{S_0, S_{1c}, S_{1s}\}$, the scattering field can be written in the global Cartesian coordinate system by employing Poisson summation formula \cite{MorseFeshbachBook}: 
\begin{eqnarray}
H_{\rm scat}(x,y) \cong e^{-\ii k_0y\sin\theta}\sum_{m=-\infty}^{+\infty}e^{-\ii m\pi y/b}\frac{e^{-\kappa_m|x|/b}}{\kappa_m}\nonumber\\
\cdot\left[\ii S_0+\frac{\ii  \kappa_mS_{1c}\sgn(x)-S_{1s}(m\pi+k_0b\sin\theta)}{k_0b}\right],
\label{ScatteringFieldGlobalCartesian}
\end{eqnarray}
where $\kappa_m=\sqrt{(m\pi+k_0b\sin\theta)^2-(k_0b)^2}$. The derivation of \eqref{ScatteringFieldGlobalCartesian} from \eqref{ScatteringFieldGlobalPolar} is also well-known \cite{TagayPaper}.

If we are interested for the electromagnetic response far from the structure ($x\rightarrow \pm\infty$), we will take into account only the refraction orders $m\in\mathbb{Z}$ belonging to the set of integers $\mathcal{M}$ defined by the double inequality \cite{PeriodicityMatters}:
\begin{eqnarray}
-{\left\lfloor \frac{2b}{\lambda}(1+\sin\theta) \right\rfloor}\le m \le {\left\lfloor \frac{2b}{\lambda}(1-\sin\theta) \right\rfloor},
\label{PropagatingRefractionOrders}
\end{eqnarray}
for which $\kappa_m$ is purely imaginary and contributes with a propagating mode to the sum \eqref{ScatteringFieldGlobalCartesian}. Obviously, if  $\frac{2b}{\lambda}<\frac{1}{1+|\sin\theta|}$, only Snell's transmission is supported by our metasurface. The scattered far-field by the grating towards the right ($x>0$) and the left ($x<0$) side can be evaluated directly from \eqref{ScatteringFieldGlobalCartesian}. In the (always present) ordinary ($m=0$) transmitted order, one should add \cite{MyJ58} the background incident field $H_{\rm inc}=e^{-\ii k_0(x \cos\theta+y \sin\theta)}$. In this sense, the transmission $T_m$ and reflection $R_m$ coefficients for $m\in\mathcal{M}$ are determined. Hence, the transmitted magnetic field reads: $H_{\rm tran}(x,y)=\sum_{m\in \mathcal{M}}T_m e^{-\ii k_0(x \cos\theta_m+y \sin\theta_m)}$ while the reflected magnetic field is equal to: $H_{\rm ref}(x,y)=\sum_{m\in \mathcal{M}}R_m e^{+\ii k_0(x \cos\theta_m-y \sin\theta_m)}$. The angle of propagation (measured from horizontal $x$ axis) of the transmitted (traveling towards $x>0$) wave of order $m$, is given by:
\begin{eqnarray}
\theta_m=\arcsin\left(\sin\theta+\frac{m\pi}{k_0b}\right).
\label{RefractionAngles}
\end{eqnarray}
If $\theta_m<0$ when $\theta>0$ for a specific $m\in\mathcal{M}$, that order of refraction is called anomalous, namely the developed transmitted rays propagate towards the same region of the space that the metasurface is excited ($y<0$ half space). For $m=0$, \eqref{RefractionAngles} yields to trivial result: $\theta_0=\theta$ (Snell's transmission).

\subsection{Nonreciprocal Diffractive Metasurfaces}
\label{sec:NonrecMetas}
The structure of Fig. \ref{fig:Fig1}(b) is designed to work as a nonreciprocal metasurface. Hence, the transmissivity of the $m$-th diffraction order, namely the portion \cite{PhysRevResearch.2.033526} of incoming power channeled along the direction $\theta=\theta_m$ from \eqref{RefractionAngles} at the half space $x>0$, reads: $\tau_m(\theta)=|T_m(\theta)|^2\cos\theta_m/\cos\theta$. Note that $T_m$ is the respective transmission coefficient for $m\in\mathcal{M}$, as defined in \eqref{PropagatingRefractionOrders}. Analogously, the reflectivity of the $m$-th diffraction order concerning the reflections at $x<0$ is written as: $\rho_m(\theta)=|R_m(\theta)|^2\cos\theta_m/\cos\theta$. In the case of a lossless system, the sum of all transmissivities and reflectivities should be equal to unity: $\sum_{m\in \mathcal{M}}\left[\tau_m(\theta)+\rho_m(\theta)\right]=1$. Having stated that, a suitable metric describing how nonreciprocally the metasurface behaves, will be the difference between the sum of transmitted and reflected power, along the horizontal $x$ axis, once the grating gets excited from different sides. In particular:
\begin{eqnarray}
\Delta P=\sum_{m\in \mathcal{M}} 
\left[\begin{array}{c}
 ~~~\left|\tau_m(\theta)-\tau_{-m}(-\theta)\right| \\
+\left|\rho_m(\theta)-\rho_{-m}(-\theta)\right|
      \end{array}\right]\cos\theta,
\label{DeltaPower}
\end{eqnarray}
where $\mathcal{M}$ is the set of propagating reflective/transmissive orders $m$ corresponding to $\theta>0$, as dictated by \eqref{PropagatingRefractionOrders}. It is important to stress that one obtains the same field distribution either the side of illumination is flipped (incident wave at the same angle $\theta$ coming from $x>0$ instead of $x<0$) or it gets excited at an opposite angle $(-\theta)$ but from the same side ($x<0$); therefore, $\Delta P$ is well-defined by \eqref{DeltaPower}. Needless to say that in the case of negative incidence angle $\theta<0$, the propagating diffractive orders $m\in\mathcal{M}$ will be the opposite ones, according to the inequality \eqref{PropagatingRefractionOrders}. Similarly, the identical effect is recorded if we keep the feeding side ($x<0$) and the angle $\theta$ the same and we just flip the sign of the magnetic bias, namely, we use a negative cyclotron frequency $\w_c<0$.

\begin{figure}[ht!]
\centering
%\subfigure[]{\includegraphics[width=4.2cm]{Fig5a}
%   \label{fig:Fig5a}}
%\subfigure[]{\includegraphics[width=4.2cm]{Fig5b}
%   \label{fig:Fig5b}}\\
%\subfigure[]{\includegraphics[width=4.2cm]{Fig5c}
%   \label{fig:Fig5c}}
%\subfigure[]{\includegraphics[width=4.2cm]{Fig5d}
   %\label{fig:Fig5d}}
   \includegraphics[width=8.4cm]{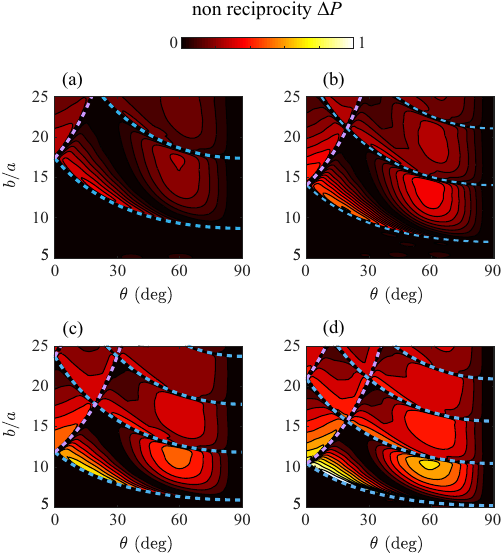}
\caption{The nonreciprocity metric \eqref{DeltaPower} of the diffractive metasurface comprising of the optimal particles of: (a) Fig. \ref{fig:Fig4}(a), (b) Fig. \ref{fig:Fig4}(b), (c) Fig. \ref{fig:Fig4}(c), (d) Fig. \ref{fig:Fig4}(d), with respect to the incidence angle $\theta$ and the normalized period $b/a$. The blue dashed lines correspond to points generating negative diffraction orders ($m<0$) and the purple dashed ones are the parametric boundaries of positive diffraction orders ($m>0$).}
\label{fig:Fig5}
\end{figure}

In an attempt to connect the scattering by isolated gyrotropic particles of Fig.~\ref{fig:Fig1}(a) with their collective operation in the infinite grating setup of Fig.~\ref{fig:Fig1}(b), we produced Fig.~ \ref{fig:Fig5}. The most nonreciprocal particles are taken from Fig.~\ref{fig:Fig2} and get analyzed in Figs \ref{fig:Fig3} and ~\ref{fig:Fig4} when placed in a metasurface configuration. In Fig. ~\ref{fig:Fig5}, we represent the collective nonreciprocity index $\Delta P$ from \eqref{DeltaPower} as a function of the incidence angle $\theta$ and the half-period of the infinite structure of Fig.~\ref{fig:Fig1}(b) normalized by the radius of the rods $b/a$. In other words, we show the variation of a quantity with respect to parameters that all have a meaning only for the collective layout instead of the individual particle. The blue dashed lines correspond to points beyond which (towards larger $b/a$) negative diffraction orders ($m<0$) are generated and, similarly, the purple dashed ones are the parametric boundaries for the onset of positive diffraction orders ($m>0$).

In Fig.~\ref{fig:Fig5}(a), we employ the maximally nonreciprocal cylinder with $a=0.04\lambda_p$ with response shown in Fig. \ref{fig:Fig4}(a). It is noticed that the nonreciprocity is overall weak due to the tiny optical size of the rods. Interestingly, $\Delta P$ almost vanishes when only Snell's channels are active since the $m=0$ diffraction order is determined chiefly from the homogenized structure. To put it alternatively, the cylinders are too tightly located and interact only via their own omnidirectional scattering mechanism (term $u=0$, coefficient $S_0$), which is reciprocal. In Fig. \ref{fig:Fig4}(b), we examine the rod of Fig. \ref{fig:Fig4}(b) with $a=0.05\lambda_p$ and realize that the nonreciprocity gets, on average, boosted. Notably, the peaks of nonreciprocal response appear very close to the parametric boundaries indicating the generation of new diffractive orders. Indeed, when a new order is born, its reflective and transmissive angles are exactly parallel to the metasurface ($y$ axis); at this precise geometric condition \cite{Hessel65}, the wave transitions from an evanescent wave (bound to the surface) to a propagating wave (radiating into free space). Since the cylinders are gyrotropic, the material introduces an asymmetric phase shift depending on the direction of wave propagation (traveling at at angles $\pm\theta$; this effectively means that the two waves experience slightly different grid periods and produce hugely different response (one evanescent, th other propagating), leading to a maximization of $\Delta P$.

In Fig.~\ref{fig:Fig5}(c), we investigate even larger rods as those of Fig.~\ref{fig:Fig4}(c) ($a=0.06\lambda_p$) and the trend of increasing the maximum nonlinearity continues. Once again, $\Delta P\rightarrow 0$ when only Snell's reflection and transmission occur while $\Delta P$ peaks are exhibited close to the aforementioned condition also known as Wood's anomaly for optical gratings \cite{Hessel65}. In Fig. \ref{fig:Fig5}(d), we utilize the largest rods with $a=0.07\lambda_0$ to build our metasurface and the same findings are recorded with the significant difference that $\Delta P$ values are very high, close to unity. Hence, we can single out two designs from that map to explore further: (i) one with close-to-normal excitation appearing just when the $m=-1$ diffractive order emerges ($\theta=17^{\circ}$, $b=8.8a$) and (ii) one working under quite oblique illumination and being very close to the onset of the $m=-2$ diffractive order ($\theta=56^{\circ}$, $b=10.5a$). We deliberately avoid to select the metasurfaces corresponding to the actual peaks, positioned very close to the dashed lines since, due to the transition from evanescent to propagating waves, our overall approximate approach (as described by, \eqref{ScatteringFieldLocalPolar},\eqref{ScatteringFieldGlobalPolar}) faces numerical issues.

\subsection{Fields Distribution in Spectrum and Space}
\label{sec:FieldSpatDist}
\begin{figure}[ht!]
\centering
%\subfigure[]{\includegraphics[width=4.2cm]{Fig6a}
 %  \label{fig:Fig6a}}
%\subfigure[]{\includegraphics[width=4.2cm]{Fig6b}
%   \label{fig:Fig6b}}
\includegraphics[width=8.6cm]{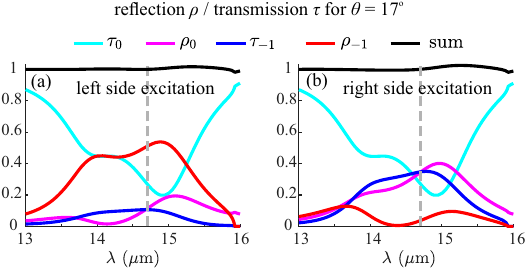}
\caption{The transmissive $\{\tau_0,\tau_{-1}\}$ and reflective $\{\rho_0,\rho_{-1}\}$ powers as functions of the operational wavelength $\lambda$ for: (a) left side excitation at $\theta=17^{\circ}$ and (b) right side excitation at $\theta=17^{\circ}$. Plot parameters: $\lambda_p=10~\mu \textrm{m}$, $a=0.07\lambda_p$, $b=8.8a$. The vertical green dashed line indicates the wavelength at which Fig.~\ref{fig:Fig5}(d) is referred to: $\lambda\cong 1.47\lambda_p\Rightarrow \w\cong 0.68\w_p$. The solid black line corresponds to the sum of all powers which, in the lossless scenario, gives unity.}
\label{fig:Fig6}
\end{figure}

With reference to the two selected designs of $a=0.07\lambda_0$ exhibiting large nonreciprocity as they correspond to the maxima of Fig.~\ref{fig:Fig5}(d), we can examine their response at a frequency spectrum around their optimal regimes. In Fig.~\ref{fig:Fig6}, we investigate the periodic array of the first selected rod. In particular, we represent the transmitted $\tau_m$ and reflected $\rho_m$ powers both for the Snell channel ($m=0$) and the additional diffractive channel ($m=-1$), as functions of the free-space wavelength $\lambda$ when the plasma wavelength is taken equal to $\lambda_p=10~\mu {\textrm m}$. Since the only propagating orders are the ones with $m=0,-1$ and the particles are lossless, we should obtain from conservation of energy: $\tau_0+\rho_0+\tau_{-1}+\rho_{-1}=1$; thus, we represent that sum with a solid black line. Note that minor discrepancies may occur since our approach is approximate and ignores rapid azimuthal variations; therefore, the obtained results rather validate our methodology. The vertical green dashed line is used to indicate the optimal regime, namely, the oscillation wavelength $\lambda\cong 1.47\lambda_p\Rightarrow \w\cong 0.68\w_p$. 

In Fig.~\ref{fig:Fig6}(a) we consider an illumination angle $\theta=17^{\circ}$ and observe that Snell transmissivity $\tau_0$ drops at the point of operation to become slightly smaller than the respective reflectivity $\tau_{-1}$. In addition, $\tau_{-1}$ tends to give a peak when nonreciprocity get substantial while $\rho_{-1}$ remains weak throughout the considered band. In Fig.~\ref{fig:Fig6}(b), we repeat our calculations for excitation at $\theta=17^{\circ}$ from the opposite side to obtain a totally different picture except for the variation for $\tau_0=\tau_0(\lambda)$ which is identical to that of Fig.~\ref{fig:Fig6}(a). In fact, one notices that we have a dominant reflection $\rho_{-1}$ along the anomalous direction $\theta_{-1}\cong 64^{\circ}$ compared to the moderate Snell transmission $\tau_0$. When it comes to the other two responses, they are relatively suppressed across the considered spectrum: at the point of operation we record $\rho_0\cong\tau_{-1}\cong 0.1$. The dissimilarity between the graphs of Fig.~\ref{fig:Fig6}(b) and Fig.~\ref{fig:Fig6}(a) demonstrate the significantly nonreciprocal behavior of the corresponding metasurface that comprises these specific rods as meta-atoms. Finally, strong nonreciprocity emerges at all the regarded oscillating wavelengths $\lambda$, which can be inferred by the wide resonances of $\Delta P$ in Fig.~\ref{fig:Fig5}(d).  

\begin{figure}[ht!]
\centering
%\subfigure[]{\includegraphics[width=7cm]{FigYa}
%   \label{fig:FigYa}}\\
%\subfigure[]{\includegraphics[width=7cm]{FigYb}
%  \label{fig:FigYass}}\\

\includegraphics[width=7.0cm]{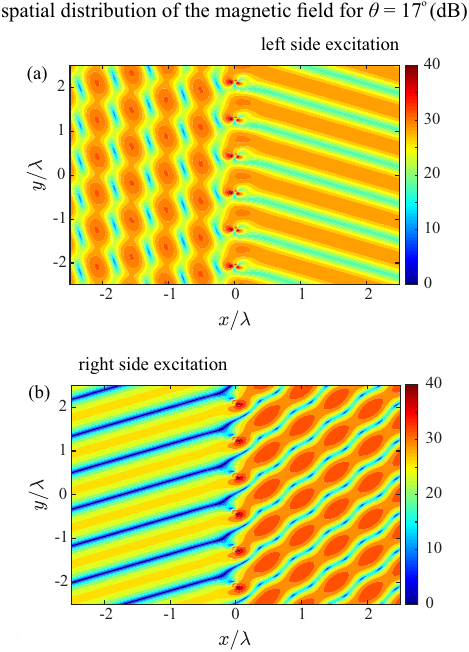}
\caption{Spatial distribution of the magnetic field relative magnitude $|H(x,y)|/|H_{\rm inc}(x,y)|$ (in decibels) of the metasurfaces corresponding to the green dashed lines of Fig.~\ref{fig:Fig6} under: (a) left side excitation at $\theta=17^{\circ}$ and (b) right side excitation at $\theta=17^{\circ}$ (Fig.~\ref{fig:Fig6}(b)). Plot parameters: $\lambda_p=10~\mu \textrm{m}$, $a=0.07\lambda_p$, $b=8.8a$, $\lambda\cong 1.47\lambda_p\Rightarrow \w\cong 0.68\w_p$.}
\label{fig:FigY}
\end{figure}

In Fig.~\ref{fig:FigY}, we depict the simulated spatial distribution of the relative magnitude of the magnetic field $|H(x,y)|/|H_{\rm inc}|$ in decibels, where $|H_{\rm inc}|=1~\textrm{A/m}$ for the metasurface of Fig.~\ref{fig:Fig6} when excited from the left (Fig.~\ref{fig:FigY}(a)) and from the right (Fig.~\ref{fig:FigY}(b)) side. The incidence angle is kept in both cases fixed at $\theta=17^{\circ}$ and the wavelength is that indicated by the green dashed lines in Fig.~\ref{fig:Fig6}: $\lambda\cong 14.7~\mu {\textrm m}$. In Fig.~\ref{fig:FigY}(a), the incident beam comes from the bottom left corner of the configuration and one may notice a certain amount of reflections as a result of the interference of three different waves: incident traveling at angle $\theta$, reflected at Snell channel $\theta_0=\theta=17^{\circ}$ and reflected along (anomalous) direction $\theta_{-1}\cong-64^{\circ}$. As far as the region with $x>0$ is concerned, substantial signal power is developed from the combination of transmission via Snell's ($m=0$) and $m=-1$ rays. Interestingly, the highest field magnitudes appear around and at the cross sections of the gyrotropic rods since only then such powerful anomalous diffraction becomes feasible. 

In Fig.~\ref{fig:FigY}(b), the response is totally different compared to that of Fig.~\ref{fig:FigY}(a) which is a heavy evidence of how nonreciprocal the structure is. Importantly, the overall reflections are higher compared to those in Fig.~\ref{fig:FigY}(a), as indicated by comparison of Figs~\ref{fig:Fig6}(a) and \ref{fig:Fig6}(b) when $\lambda\cong 14.7~\mu {\textrm m}$. On the other hand, transmission is very weak as again shown in Fig.~\ref{fig:Fig6}(b). Notice that we represent the magnitude of the field; hence the one-dimensional oscillations in the transmissive area are owed to the interaction of two waves (traveling along angles $\theta$ and $\theta_{-1}$, towards the same direction of axis $x$), not to the diffusion of $\Re[H(x,y)]$ for a single wave. With regard to the half space $x>0$, the emerged two-dimensional perturbations are the outcome of the coexistence of three waves (incoming at angle $\theta$, Snell refractive at angle $(-\theta)$ and anomalous refractive at angle $\theta_{-1}$), two of which propagate towards opposite $x$ directions.

\begin{figure}[ht!]
\centering
%\subfigure[]{\includegraphics[width=4.2cm]{Fig7a}
%   \label{fig:Fig7a}}
%\subfigure[]{\includegraphics[width=4.2cm]{Fig7b}
%   \label{fig:Fig7b}}
\includegraphics[width=8.4cm]{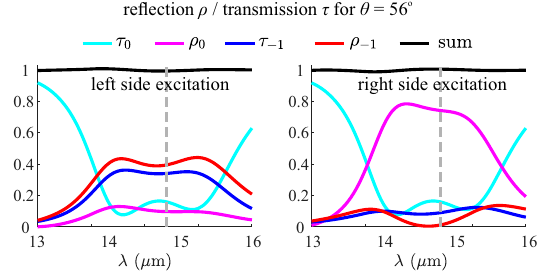}
\caption{The transmissive $\{\tau_0,\tau_{-1}\}$ and reflective $\{\rho_0,\rho_{-1}\}$ powers as functions of the operational wavelength $\lambda$ for: (a) left side excitation at $\theta=56^{\circ}$ (Fig.~\ref{fig:Fig7}(a)), (b) right side excitation at $\theta=56^{\circ}$. Plot parameters: $\lambda_p=10~\mu \textrm{m}$, $a=0.07\lambda_p$, $b=10.5a$. The vertical green dashed line indicates the wavelength at which Fig.~\ref{fig:Fig5}(d) is referred to: $\lambda\cong 1.47\lambda_p\Rightarrow \w\cong 0.68\w_p$. The solid black line corresponds to the sum of all powers which, in the lossless scenario, gives unity.}
\label{fig:Fig7}
\end{figure}

In Fig.~\ref{fig:Fig7}, we consider a periodic grating consisting the second selected rod from Fig.~\ref{fig:Fig5}(d) as a module and redo the evaluations of Fig.~\ref{fig:Fig6}. In Fig.~\ref{fig:Fig7}(a), we represent the transmissivities $\{\tau_0,\tau_{-1}\}$ and reflectivities $\{\rho_0,\rho_{-1}\}$ of the two guided diffraction orders with respect to operational wavelength $\lambda$ for left-side incidence at the angle $\theta=56^{\circ}$ and, again, with a gyrotropic medium of $\lambda_p=10~\mu {\textrm m}$. One directly observes that responses at $\theta_{-1}\cong -5^{\circ}$ are very strong towards both half spaces engulfing our metasurface ($\tau_{-1}$ and $\rho_{-1}$) while Snell channels carry small portions of power. On the contrary, when the same structure is illuminated from the right side (Fig.~\ref{fig:Fig7}(b)), we record a very pronounced Snell's reflection while all the rest diffractive powers stay relatively suppressed. As expected, the metasurface exhibits a significant directional dependence in the distribution of power among the available diffraction channels. While the total scattered power remains conserved (a fact that again validates the followed approximations), the relative efficiencies of the transmitted and reflected diffraction orders differ significantly for the two incidence directions. Such a major asymmetry at a larger incidence angle ($\theta=56^{\circ}$) confirms that the observed behavior is an intrinsic consequence of the gyrotropic response of the constituent rods rather than a particular feature associated with a specific excitation condition. 

\begin{figure}[ht!]
\centering
%\subfigure[]{\includegraphics[width=6.2cm]{FigZa}
%   \label{fig:FigZa}}\\
%\subfigure[]{\includegraphics[width=6.2cm]{FigZb}
%   \label{fig:FigZb}}
\includegraphics[width=7.2cm]{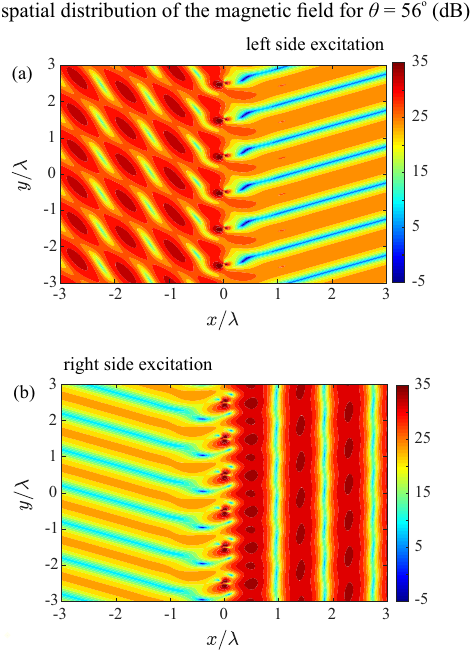}
\caption{Spatial distribution of the magnetic field relative magnitude $|H(x,y)|/|H_{\rm inc}(x,y)|$ (in decibels) of the metasurfaces corresponding to the green dashed lines of Fig.~\ref{fig:Fig7} under: (a) left side excitation at $\theta=56^{\circ}$ (Fig.~\ref{fig:Fig7}(a)), (b) right side excitation at $\theta=56^{\circ}$ (Fig. \ref{fig:Fig7}(b)). Plot parameters: $\lambda_p=10~\mu \textrm{m}$, $a=0.07\lambda_p$, $b=10.5a$, $\lambda\cong 1.47\lambda_p\Rightarrow \w\cong 0.68\w_p$.}
\label{fig:FigZ}
\end{figure}

In Fig.~\ref{fig:FigZ}, we operate the structure investigated in Fig.~\ref{fig:Fig7} at exactly the wavelength $\lambda\cong 14.7 \mu {\textrm m}$ and  an illumination angle $\theta=56^{\circ}$ giving substantially nonreciprocal behavior, according to the indicator of Fig.~\ref{fig:Fig5}(d) and the comparison of Fig.~\ref{fig:Fig7}. In the simulation of Fig.~\ref{fig:FigZ}(a), we observe stronger fields in the reflection space ($x<0$) rather than in the transmission ($x>0$) one, as well-predicted by the curves of Fig.~\ref{fig:Fig7}(a). Interestingly, the maximum fluctuation of the signal at $x<0$ is not occurring along the horizontal axis as happens when the Snell reflection is strong and close to the incident amplitude but along an oblique direction revealing the substantial values of $\rho_{-1}$. In Fig.~\ref{fig:FigZ}(b), we feed the structure from the right port and notice very powerful reflections at $x>0$; indeed, inspection of the spatial variations say that $\rho_0$ is so strong that almost creates standing waves. To put it mathematically, $|R_0|$ is so big that gives: $\left|e^{-\ii k_0 x \cos\theta}+R_0 e^{+\ii k_0 x \cos\theta}\right|^2 \cong 2\left\{1+\cos(k_0 x \cos\theta+\textrm{Arg}[R_0])\right\}$. When it comes to the transmissive signal at $x<0$, it is much weaker than that of Fig.~\ref{fig:FigY}(a) (for $x>0$); such a feature is another manifestation for the heavily nonreciprocal response of selected metasurface. In addition, the difference between the two transmitted powers can be understood from Fig.~\ref{fig:Fig7}(b); indeed, at $\lambda\cong 14.7~\mu {\textrm m}$, the Snell transmissivity $\tau_0$ is again equal to that of Fig.~\ref{fig:Fig7}(a) but the quantity $\tau_{-1}$ is much smaller compared to the respective one of Fig.~\ref{fig:Fig7}(a).   

\section{Conclusions}
\label{sec:SectionIV}
Subwavelength plasmonic cylinders under an axial static magnetic field exhibit a remarkably strong, nonreciprocal electromagnetic response. Formulating an intuitive metric based on cylindrical wave expansion allows for the precise identification of operational frequencies and bias conditions that maximize this nonreciprocal behavior, yielding highly asymmetric scattering patterns that can be dynamically inverted by reversing the DC field sign. A multipolar decomposition reveals that this extreme nonreciprocity is physically rooted in a balanced magneto-electric interaction, driven primarily by the overlapping contributions of electric and magnetic dipolar modes. When translated from isolated particles into a periodic array, these optimized cylinders form a macroscopic metasurface characterized by stark, direction-dependent transmission and reflection under opposite oblique incidences. The maximum collective nonreciprocity aligns directly with the emergent boundaries of new diffraction orders, opening new avenues for anomalous wavefront engineering.

Looking forward, several promising directions emerge to expand upon the foundations established in this study. In particular, transitioning from homogeneous to multilayered gyrotropic rods \cite{ Valagiannopoulos2012-ga}, such as core-shell configurations, could offer additional degrees of design freedom to fine tune the magneto-electric overlap and further enhance nonreciprocal scattering. In addition, investigating finite clusters \cite{Valagiannopoulos2011-tt,PhysRevB.102.155310} of these gyrorods, rather than infinite periodic arrays, will provide crucial insights into localized near field coupling and the design of compact nonreciprocal photonic components. Interestingly, substituting circular cylinders with elliptic rod metasurfaces \cite{PhysRevB.94.205433} will introduce geometric anisotropy, an avenue that can be exploited to break spatial symmetries alongside time-reversal symmetry for more sophisticated wavefront manipulation. Finally, exploring the interplay between magnetic bias and intrinsic optical activity within chiral metasurfaces \cite{ Katsantonis2023-wt} could unlock novel nonreciprocal phenomena, paving the way for advanced chiroptical devices and topological photonic phases.

\begin{acknowledgments}
This work was supported by the National Recovery and Resilience Plan Greece 2.0, funded by the European Union – NextGenerationEU (Implementation body: HFRI), Project No. 14830, PhoToCon and by the European Union under the call “EIC Pathfinder Open 2022” (Project PULSE, Project No. 101099313).
\end{acknowledgments}

\bibliography{TOPOGRA_Bibliography}

\end{document}